\documentclass[11pt]{iopart}
\usepackage{amssymb}
\usepackage{amsfonts}
\usepackage{hyperref}

\renewcommand{\d}{\mathrm{d}}
\newcommand{\pidS}{\pi_{\mathrm{dS}}}
\renewcommand{\etal}{\emph{et al.}}
\newcommand{\Mp}{M_{\mathrm{P}}}
\renewcommand{\geq}{\geqslant}

\newcommand{\Ricci}{\mathcal{R}}
\newcommand{\RSch}{R_{\mathrm{S}}}
\newcommand{\Rbound}{R_{\mathrm{bound}}}

\begin{document}

	\begin{flushright}
		\textsf{DESY 10-062}
	\end{flushright}
	\title{Revisiting fifth forces in the Galileon model}
	
	\author{Clare Burrage$^1$ and  David Seery$^2$ }
	\vspace{3mm}
	
	\address{$^1$ Theory Group,
			Deutsches Elektronen-Synchrotron DESY, D-22603, \\
			Hamburg, Germany \\[2mm]
		$^2$ Department of Physics and Astronomy, University of
        	Sussex, \\ Brighton, BN1 9QH, UK}
	\vspace{3mm}

	\eads{\mailto{clare.burrage@desy.de}, \mailto{D.Seery@sussex.ac.uk}}
	
	\begin{abstract}
		A Galileon field is one which obeys a spacetime generalization of
		the non-relativistic Galilean invariance.
		Such a field may possess non-canonical kinetic terms,
		but ghost-free theories with a well-defined Cauchy problem
		exist,
		constructed using a finite number of relevant operators.
		The interactions of this scalar with matter are hidden by the
		Vainshtein effect, causing the Galileon to become weakly coupled
		near heavy sources. We revisit estimates
		of the fifth force mediated by
		a Galileon field, and show that the parameters of the model are
		less constrained by experiment that previously supposed.                
	\end{abstract}
	\maketitle

\section{Introduction}
\label{sec:intro}

Many disparate cosmological observations suggest that
the universe recently entered an accelerating epoch.
The simplest
explanation for this acceleration is the inclusion of a cosmological
constant $\Lambda \sim (10^{-3}\;\mbox{eV})^4$ in the Einstein equations.
Within classical general relativity there is no reason to prefer one value
of $\Lambda$ over another. The inclusion of particle
physics, however, leads to a natural interpretation of $\Lambda$ as
the energy density of the vacuum.  Field-theoretic estimates of this
energy density diverge as the fourth power of the cutoff, $M$,
yielding $\Lambda \sim M^4$.
A realistic choice almost certainly requires
$M \gg \mbox{TeV}$,
which markedly disagrees with the tiny value measured by experiment.%
	\footnote{If the cutoff for field
		theory is assumed to be the Planck scale, there is a mismatch
		between the predicted and observed values of $\Lambda$ by
		approximately 120
		orders of magnitude.}

The ``old'' cosmological constant problem is the question:
why is the observed value of the cosmological constant so small
\cite{Weinberg:2000yb}?
But we can just as well ask:
why does the vacuum energy gravitate so little \cite{Dvali:2007kt}?
If long wavelength fluctuations gravitate more weakly than other sources,
the vacuum energy may be screened or ``degravitated'' \cite{Dvali:2002pe,ArkaniHamed:2002fu} in the same way that
long wavelength excitations far beyond the Debye sphere are screened
by the effective photon mass in a plasma.
By analogy,
degravitation occurs if the graviton can acquire a mass,
perhaps because it occurs as a resonance \cite{Dvali:2007kt}
or extended object \cite{Sundrum:2003jq},
or because it descends from a braneworld scenario such as
the DGP model or its generalizations
\cite{deRham:2007xp,deRham:2007rw}.
An implementation of Higgs' mechanism for the graviton has recently
been constructed by Chamseddine \& Mukhanov \cite{Chamseddine:2010ub}.

The four-dimensional effective theory of such models
necessarily includes extra scalar fields, arising from new longitudinal
polarizations of the graviton.
It is these new modes
which are the agent of degravitation,
requiring them to couple to gravitational sources.
Therefore one may have some reservations that these scalars mediate
unacceptably large fifth forces.
However, it is possible for these
fields to avoid experimental constraints using the
Vainshtein (``kinetic chameleon'')
mechanism \cite{Vainshtein:1972sx,Deffayet:2001uk}, which exploits their
non-canonical kinetic terms.
Non-linearities become important
in the vicinity of dense objects, and the
scalars naturally become weakly coupled to matter.
A successful Vainshtein decoupling may require
less delicate fine-tuning of parameters.

Nicolis, Rattazzi \& Trincherini \cite{Nicolis:2008in}
discussed the most general
four-dimensional ghost-free Lagrangian for a scalar field with these
properties. Their construction invoked a gradient
shift symmetry for the scalar field, a generalization of non-relativistic
Galilei invariance, which permits sizeable non-linearities to occur.
Nicolis {\etal} noticed that a finite set of operators were consistent
with this symmetry and simultaneously gave rise to second-order
equations of motion.
Taking these operators as the relevant terms in an effective
field theory, with all other operators considered as
irrelevant deformations, one arrives at a perturbatively unitary
Lagrangian \cite{Donoghue:1994dn,Donoghue:1993eb}.
Moreover,
the relevant operators are protected against renormalization from
scalar loops
\cite{Luty:2003vm,Porrati:2004yi}.

Nicolis {\etal} denoted the field described by the resulting Lagrangian
as $\pi$, and referred to it as a \emph{Galileon}.
Couplings of the Galileon to gravity
were studied by Deffayet {\etal} \cite{Deffayet:2009wt,Deffayet:2009mn},
and 
aspects of its cosmological phenomenology were explored
by numerous authors
\cite{Chow:2009fm,Silva:2009km,Kobayashi:2010wa,Kobayashi:2009wr,Gannouji:2010au}. The gravitational backreaction of the Galileon
field was shown to be small in Ref.~\cite{VanAcoleyen:2009kb}.
An explicit higher dimensional theory, for which the Galileon is the
four-dimensional effective theory, was discussed
in Ref.~\cite{deRham:2010eu}.

The Galileon Lagrangian contains at least five dimensionful parameters,
one for each of the relevant operators.%
	\footnote{Other parameters may be associated with irrelevant deformations,
		which we ignore in the present discussion.}
These may be chosen freely subject to certain constraints.
Of these, the most important is that
the acceleration of the Hubble rate should agree
with our observations.
The remaining parameters must combine in such a way that
fifth forces mediated by the Galileon are phenomenologically acceptable.
These constraints were discussed by Nicolis {\etal}
\cite{Nicolis:2008in}.
In this paper we revisit this analysis
and show that the Vainshtein radius of a massive
object---within which the equations of motion of the Galileon become
non-linear and the scalar force it mediates is dynamically suppressed---%
can be made arbitrarily large by varying a single combinations
of the parameters.  The magnitude of the scalar force within the
Vainshtein radius depends on an independent combination of
parameters.
When a sufficient number of relevant operators are included, these quantities
may therefore vary independently.

This freedom leaves the parameters of the Galileon model weakly
constrained by gravitational experiments.  With this in mind we revisit
the strong-coupling limit of the theory and show that,
in contrast to the DGP
scenario, the distance scale at which
non-renormalizable corrections become important
may be far below the reach of current experimental probes.

This brief note is organized as follows.
In \S\ref{sec:eft} we review the action for a  Galileon
scalar field.  In \S\ref{sec:fifth} the fifth forces mediated by Galileon
exchange are calculated, and we demonstrate that the constraints on
the parameters of the theory are not as stringent as previously
supposed.  For this broader parameter space, the scale at which quantum
fluctuations become strongly interacting is derived in
\S\ref{sec:strong}. We conclude in \S\ref{sec:conc}. In
an Appendix we give a simple
derivation of the Galileon force law assumed throughout this article.
We choose units in which $c = \hbar = 1$.

\section{Galileon effective field theories}
\label{sec:eft}

It was explained in \S\ref{sec:intro} that Galileon theories are
associated with modifications of gravity in the infrared,
owing to the presence of a scalar field, $\pi$, which mixes kinetically
with the metric.
We are free to choose the precise coupling of $\pi$ to matter,
for which
various possibilities were discussed by de Rham \& Tolley
\cite{deRham:2010eu}.
In Ref.~\cite{Nicolis:2008in} the $\pi$-action was assumed to take the
form
\begin{equation}
	S_{\pi} = \int \d^4 x \; \Big(
		\mathcal{L}_\pi + \pi {T^\mu}_\mu
	\Big) ,
	\label{eq:action}
\end{equation}
where $T_{\mu\nu}$ is the energy--momentum tensor of matter,
which couples to the perturbative graviton $h_{\mu\nu}$ as expected,
via $h_{\mu \nu} T^{\mu \nu}/2$.
We allow the self-interactions of $\pi$ to include
quite general non-linearities, but
treat the coupling between $\pi$ and other fields linearly.
In addition, we neglect back-reaction on the metric
of any energy density associated
with $\pi$. These simplifications are standard in the
literature.

The Galilean shift symmetry imposed by Nicolis {\etal} was
\begin{equation}
\pi(x) \rightarrow \pi(x) +b_{\mu}x^{\mu}+c ,
\end{equation}
where $b_\mu$ and $c$ are constants. In addition
to invariance under this symmetry, as discussed above,
the relevant operators in the effective theory for $\pi$ must be chosen
to yield second-order equations of motion. There are five such operators
\cite{Nicolis:2008in}
\begin{eqnarray}
\mathcal{L}_1&=&\pi\\
\mathcal{L}_2&=&-\frac{1}{2}\partial\pi\cdot\partial\pi\\
\mathcal{L}_3&=&-\frac{1}{2}(\Box\pi)\partial\pi\cdot\partial\pi\\
\mathcal{L}_4&=&-\frac{1}{4}[(\Box\pi)^2\partial\pi\cdot\partial\pi-2(\Box\pi)\partial
\pi \cdot \Pi\cdot\partial\pi\nonumber\\
 & &\;\;\;\;\;\;\;\;\; -
(\Pi\cdot\Pi)(\partial\pi\cdot\partial\pi)+2\partial\pi\cdot\Pi\cdot\Pi\cdot\partial\pi]\\
\mathcal{L}_5&=&-\frac{1}{5}[(\Box\pi)^3\partial\pi\cdot\partial\pi
-3(\Box\pi)^2\partial\pi\cdot\Pi\cdot\partial\pi-3\Box\pi(\Pi\cdot\Pi)(\partial\pi\cdot\partial\pi)\nonumber\\
& &\;\;\;\;\;\;\;\;\;
+6(\Box\pi)\partial\pi\cdot\Pi\cdot\Pi\cdot\partial\pi
+2(\Pi\cdot\Pi\cdot\Pi)(\partial\pi\cdot\partial\pi) \nonumber\\
& &\;\;\;\;\;\;\;\;\;
+3(\Pi\cdot\Pi)\partial\pi\cdot\Pi\cdot\partial\pi -
6\partial\pi\cdot\Pi\cdot\Pi\cdot\Pi\cdot\partial \pi ]
\end{eqnarray}
where `$\cdot$' indicates contraction of spacetime indices, and
$\Pi^{\mu}_{\nu}\equiv \partial^{\mu}\partial_{\nu}\pi$.  
The relevant part of the $\pi$ effective theory must be constructed
out of these operators, making its Lagrangian of the form
\begin{equation}
\mathcal{L}_{\pi}=\sum_{i=1}^5c_i\mathcal{L}_i ,
\label{eq:lag}
\end{equation}
where the $c_i$ are arbitrary parameters.

For example,
the DGP model \cite{Dvali:2000hr}
is recovered from \eref{eq:lag} by the choices
$c_1=c_4=c_5=0$, $c_2=24 M^2_4$ and $c_3=16(M_4/ \Lambda)^3$, where $M_4$
is the four dimensional Planck mass, and $\Lambda \sim
(M_4/L_{\mathrm{DGP}})^{1/3}$ where $L_{\mathrm{DGP}}$
is the distance scale at which
gravity weakens and becomes five dimensional.  In a five-dimensional
realization of the Galileon model \cite{deRham:2010eu},
constructed from
a four-dimensional DBI probe brane embedded in a five-dimensional
spacetime, $c_2$ is related to the brane tension, and the higher
coefficients $c_3$, $c_4$ and $c_5$ receive contributions from the
normalization of higher curvature invariants in the five-dimensional
brane action.  
They do not receive perturbative renormalizations from $\pi$ loops
\cite{Luty:2003vm,Porrati:2004yi}, although we would generically expect
corrections from loops among matter living on the brane.
Therefore,
in this realization, such corrections may impose naturalness relations
among the parameters.

\section{Fifth forces from the Galileon}
\label{sec:fifth}

If the Galileon model is to give an acceptable phenomenology, it must
generate an accelerating universe at late times.
Therefore, we expect that $\pi$ has a solution
corresponding to de Sitter space,
even in the absence of sources from other matter fields.
Nicolis {\etal} argued that the necessary solution could be
written \cite{Nicolis:2008in}
\begin{equation}
	\pidS(x)=-\frac{1}{4}H^2x_{\mu}x^{\mu}
	\label{eq:deSitter}
\end{equation}
where $H$ is the Hubble parameter of the de Sitter metric.
The existence of this solution imposes an algebraic constraint
on the coefficients $c_i$,
\begin{equation}
	c_1-2c_2H^2+3c_3H^4-3c_4H^6+\frac{3}{2}c_5H^8=0 .
	\label{deSitter}
\end{equation}
In the presence of matter, the Galileon is perturbed from
its background de Sitter solution~\eref{eq:deSitter}.
It is these perturbations which may mediate a fifth force, potentially
in conflict with accurate experimental tests of gravity.

\subsection{The Vainshtein radius}
\label{sec:vainshtein}

To study solutions of the the Galileon field equations
near a massive object, we perturb about the background de Sitter
solution
\begin{equation}
	\pi\rightarrow \pidS + \pi ,
\end{equation}
in which we have adopted the notation of Nicolis {\etal}
in denoting the Galileon perturbation $\pi$. To avoid ambiguity
we use this notation exclusively for the remainder of this
article, writing $\pidS$ explicitly when we
require the background solution and $\pidS + \pi$ for the full field.

It follows from~\eref{eq:action} and~\eref{eq:deSitter}
that the equations of motion for $\pi$ are
\begin{equation}
	\sum_{i=2}^5d_i\mathcal{E}_i=-T^{\mu}_{\mu} ,
\end{equation}
where the $d_i$ are linear combinations of the $c_i$
\begin{equation}
	\left(\begin{array}{c}
		d_2\\
		d_3\\
		d_4\\
		d_5
	\end{array}\right)
	=
	\left(\begin{array}{cccc}
		1&-3H^2&\frac{9}{2}H^4&-3H^6\\
		0&1&-3H^2&3H^4\\
		0&0&1&-2H^2\\
		0&0&0&1
	\end{array}\right)
	\left(\begin{array}{c}
		c_2\\
		c_3\\
		c_4\\
		c_5
	\end{array}\right)
\label{eq:dc}
\end{equation}
and the $\mathcal{E}_i$ are functions of $\partial_\mu \partial_\nu \pi$
which
coincide with Eqs.~(39)--(43) of Ref.~\cite{Nicolis:2008in}.

Following Ref.~\cite{Nicolis:2008in},
we study the spherically symmetric, static equations of motion for the
scalar field near an object of mass $M$ localized at the origin
\begin{equation}
	d_2\left(\frac{\pi^{\prime}}{r}\right) +
	2d_3\left(\frac{\pi^{\prime}}{r}\right)^2+2d_4
	\left(\frac{\pi^{\prime}}{r}\right)^3=\frac{M}{4\pi r^3} .
	\label{eq:eom}
\end{equation}
The $\mathcal{L}_5$ operator does not contribute to this solution,
since it is identically zero when evaluated on a static, radial
profile. Therefore $d_5$ is unconstrained by this procedure.
When $d_3=d_4=0$ we recover the linear solution
\begin{equation}
	\pi(r)=\pi_0-\frac{M}{4\pi d_2r} ,
\end{equation}	
for which the ratio of Galileon to Newtonian gravitational forces is
\begin{equation}
	\frac{F_{\pi}}{F_N}=\frac{M_P^2}{4\pi d_2} .
\label{eq:linforce}
\end{equation}
In this case,
modifications of gravity due to the scalar field can be suppressed
only by making the Galileon very weakly coupled,
requiring $d_2 \gg M_P^2$, although
in the absence of an ultraviolet completion for general
relativity the meaning of masses larger than
$M_P$ is unclear.  If we adopt this strategy, it is hard to
justify stability of the resulting scenario against radiative
corrections \cite{Carroll:1998zi}.

The situation is very different if at least one of $d_3$ or $d_4$
is not zero, because non-linearities in Eq.~\eref{eq:eom}
necessarily become important.
We write
\begin{equation}
	\frac{\pi^{\prime}}{r}=\frac{M}{4\pi r^3 d_2} g(r) ,
	\label{eq:gausslaw}
\end{equation}
and define two lengthscales, associated with the strength of the
$\mathcal{L}_3$ and $\mathcal{L}_4$ operators and the mass of the
object $M$,
\begin{eqnarray}
	R_1^3&=&\frac{d_3 M}{2\pi d_2^2} ,\\
	R_2^6&=&\frac{M^2d_4}{8\pi^2d_2^3} .
\end{eqnarray}
In terms of these quantities, Eq.~\eref{eq:eom} can be given a simple
expression,
\begin{equation}
	g+\left(\frac{R_1}{r}\right)^3g^2+\left(\frac{R_2}{r}\right)^6g^3=1 .
	\label{eom}
\end{equation}
The $\pi$-profile sources a $1/r^2$ force law whenever $g = 1$.
The strength of this force is set by $d_2$, which has dimensions of
$[\mbox{GeV}^2]$, giving an effective `Galileon Newton's constant'
$G_\pi \sim 1/d_2$. This $1/r^2$
force will be stronger than
gravity---and therefore in potential conflict with observation---if
$d_2 \lesssim \Mp^2$.
However, the force can be made
substantially  weaker if $g$ becomes much
smaller than unity.

When can we expect significant suppression?
Eq.~\eref{eom} shows clearly that, whenever
$R_1$ or $R_2$ are non-zero,
the solution $g(r)$ must approach zero sufficiently rapidly
as $r \rightarrow 0$. Therefore, Eq.~\eref{eom} automatically suppresses
the $\pi$-force within a radius $r < r_\star$, where
the strong-coupling or `Vainshtein' radius
$r_\star$ is controlled by the onset of non-linear behaviour
in Eq.~\eref{eom}.
To obtain an estimate one can determine the radius $r_n$ at which
$g$ has decreased to $\sim 1/n$, where $n$ is a small natural number,
\begin{equation}
	r_n^3 = \frac{R_1^3 +\sqrt{R_1^6 +4(n-1)R_2^6}}{2n(n-1)} .
\end{equation}
If any hierarchy exists between $R_1$ and $R_2$, then
$r_n$ is approximately given by the larger of the two.
On the other hand,
if $R_1 \sim R_2$ then $r_n \sim R_1 \sim R_2$. In either case
we can make a good
estimate of the Vainshtein radius by setting
$r_\star \sim \max \{ R_1, R_2 \}$.
This remarkable, automatic suppression of the $\pi$-force
for $r < r_\star$ is the basis of the Vainshtein mechanism,
allowing fifth forces to be dynamically suppressed without fine-tuning
parameters.

It is possible to set $R_1$ and $R_2$ to arbitrary values by varying
the Lagrangian coefficients $c_1$, $c_2$ and $c_3$, which allows the
Vainshtein radius to be made arbitrarily large or small.
This is more general than Vainshtein's original scenario
\cite{Vainshtein:1972sx}.
It is also more general than the first non-trivial Galileon
scenario, which was realized within the DGP model
\cite{Dvali:2000hr}. In this case the Vainshtein radius
 for an object of mass $M$ depends only on the
Planck scale, $\Mp$, and the scale of the de Sitter cosmological
constant,
\begin{equation}
	r_\ast \sim \left( \frac{M}{4\pi} \frac{1}{\Mp^2 H_0^2} \right)^{1/3} .
\end{equation}
This happens because the DGP model fixes the Lagrangian parameters
to be combinations of the Planck and Hubble scales up to $\Or(1)$ constants.

\subsection{Suppression within the Vainshtein radius}
\label{sec:suppress}

We have seen that non-linearities inherent in the Galileon model
dynamically suppress $\pi$-exchange forces within the Vainshtein radius
of a massive object; the suppression factor $g(r)$ in
Eq.~\eref{eq:gausslaw} must tend to zero as the origin is approached.
In this section we argue that
if the `bare' coupling constant of the far-field $1/r^2$ force is too large,
so that $d_2\ll M_P^2$, dynamical suppression may not be enough to
hide the Galileon from experimental probes of gravity, which are
conducted at a finite distance from the origin.
It is necessary to apply a further restriction
to the Lagrangian parameters, controlling the
degree of suppression when $r \ll r_\star$.

We first compute the Galileon force within the Vainshtein region.
In Ref.~\cite{Nicolis:2008in}
it was argued that stable solutions to the
equations of motion exist only if $d_2>0$,
$d_4\geq 0$, $d_3\geq \sqrt{(3/2)d_2d_4}$ and $d_5<0$.  This entails
the relation $R_1 \geq 3^{1/6} R_2 > R_2$. Also, it is useful to introduce
a quantity $\alpha$ which measures the hierarchy between
$\mathcal{L}_3$ and $\mathcal{L}_4$ in terms of their associated
radii $R_1$ and $R_2$,
\begin{equation}
	\alpha \equiv \left(\frac{R_2}{R_1}\right)^3 < 1 .
\end{equation}
If both $R_1$ and $R_2$
are non-zero, there are two spatial regions in which
the solution can be approximated easily:
\begin{itemize}
\item {\bf Region A:} $\alpha R_2 \ll r \ll R_1$.

This implies $g \sim (r/R_1)^{3/2}$, meaning that the ratio
of the scalar ($F_\pi$) to gravitational ($F_N$) force is
\begin{equation}
	\frac{F_{\pi}}{F_N}
	=
	\frac{M_P^2}{4\pi d_2}\left(\frac{r}{R_1}\right)^{3/2}
	=
	\frac{M_P^2}{2}\left(\frac{1}{2\pi d_3M}\right)^{1/2}r^{3/2} .
	\label{eq:FA}
\end{equation}

\item {\bf Region B:}  $0<r \ll \alpha R_2$.

In this region $g \sim  (r/R_2)^{2}$, meaning that the ratio
of the scalar to gravitational force is
\begin{equation}
	\frac{F_{\pi}}{F_N}
	=
	\frac{M_P^2}{4\pi d_2}\left(\frac{r}{R_2}\right)^2
	=
	\frac{M_P^2}{2}\left(\frac{1}{\pi d_4 M^2}\right)^{1/3}r^2
	\label{eq:FB}
\end{equation}

\end{itemize}
Clearly Region B only exists when $d_4\neq 0$.
In principle, Region B could extend from zero to $r \sim r_\star$
if $d_3$ is absent. However, this arrangement is excluded by the
stability argument discussed above which requires $R_1 > R_2$.

Eqs.~\eref{eq:FA}--\eref{eq:FB} show clearly that for $r < r_\star$,
the Galileon force is independent of $d_2$.
This effect was noticed very recently by
Gannouji \& Sani \cite{Gannouji:2010au}.
The interpretation is simple:
the scale $d_2$ controls the strength of the Galileon force in the
linear regime, where $\pi$-exchange obeys a $1/r^2$ force law.
Therefore it is clear that $d_2 \gtrsim M_P^2$
is \emph{sufficient} to ensure Galileon forces are suppressed
within the Vainshtein radius, because whenever $r$ falls appreciably
below $r_\star$ we will find that $F_\pi / F_N$ becomes negligible.
This choice was made in Ref.~\cite{Nicolis:2008in}.

However, this is not necessary.
The linear region, where a $1/r^2$ force law prevails, may
never probed be directly by experimental tests of
gravity. Instead,
experiments typically study
the motion of a test mass in the gravitational field of a much
more massive source.
What conditions must we impose?
First,
if the Galileon is to hide dynamically from experiments seeking
modifications of gravity, then the entire experiment must lie within its own
Vainshtein radius.
Second, Eqs.~\eref{eq:FA}--\eref{eq:FB}
for the Galileon force show that, provided we know the separation of
source and test masses, and the source mass $M$,
then gravitational tests constrain either $d_3$ or $d_4$.
For consistency we must ensure that the Vainshtein radius of
the source mass is sufficiently large to contain the experimental
apparatus---requiring a tuning of the independent combinations of
Lagrangian parameters in $R_1$ and $R_2$.

These constraints do not ensure that the Galileon force is weaker than
gravity everywhere within the Vainshtein radius. Indeed,
for $r \sim r_\star$ we find $F_\pi / F_N \sim \Mp^2 / d_2$ which
may be large.
However, the freedom to make the
Vainshtein radius of any object arbitrarily large means it is never
necessary to impose the much more stringent condition
that $F_\pi / F_N \ll 1$ for all $r < r_\star$.
We can therefore extend the parameter range which was allowed in
Ref.~\cite{Nicolis:2008in} to include the region where $d_2 < \Mp^2$,
without damage to a successful Vainshtein
decoupling.
For example, in the de Rham--Tolley construction
discussed in \S\ref{sec:eft} one has $c_2 \sim \lambda$, where $\lambda$
is the tension of the 3-brane. In this construction, we conclude that
even in backgrounds where  $c_2 \sim d_2$,
the tension may be adjusted as desired, without
being constrained to the Planck scale.

\subsection{Gravitationally bound objects}

We leave detailed calculation of the constraints imposed by laboratory and
astronomical tests for future work.
However, we can obtain an estimate of the freedom allowed in $d_2$,
$d_3$ and $d_4$ by imposing a simple and generic requirement:
gravitationally bound objects should not feel significant deviations
from general relativity. 
This is necessary because
the validity of post-Newtonian gravity has been established for a wide
variety
of gravitationally bound systems. 

We denote the Ricci curvature $\Ricci$ and approximate it
near a massive object by
$\Ricci \sim \RSch /r^3$, where $\RSch$ is the Schwarzschild radius of
the object. A system is gravitationally bound if its
gravitational attraction is stronger than the Hubble flow, which requires
$\Ricci \gtrsim H_0^2$.  We can reformulate this
constraint by asking that the system
is entirely contained within the radius
\begin{equation}
	r \lesssim \Rbound \equiv \left(\frac{\RSch}{H_0^2}\right)^{1/3}
\end{equation}

According to the argument outlined in \S\ref{sec:suppress},
if the Galileon is to evade experimental probes of gravity
dynamically then we must require that $\Rbound$ is smaller than the
radius at which the gravitational and Galileon forces are equal in
magnitude
\begin{equation}
R_{\mathrm{equality}}^3 = 8\pi \frac{d_3 M}{M_P^4}.
\end{equation}
The requirement $\Rbound \lesssim  R_{\mathrm{equality}}$ therefore
imposes the constraint
\begin{equation}
d_3 \gtrsim 10^{118}.
\label{eq:constr}
\end{equation}
This can be written as a constraint on the Lagrangian parameters by
making use of Eq.~\eref{eq:dc}.  For consistency, to ensure that the
Vainshtein radius is at least as large as $R_{\mathrm{equality}}$, we
must also require $d_2\lesssim M_P^2$.

In Ref.~\cite{Nicolis:2008in} it was argued that for gravitationally
bound objects to be within their own Vainshtein radius it was
necessary to have $d_2\sim M_P^2$.  Here we have demonstrated
that, because in the Galileon model we have the freedom to vary
the Vainshtein radius of an object, this is too restrictive a constraint.

\section{The strong interaction scale}
\label{sec:strong}

In the DGP model the Lagrangian parameters $c_i$ are
constrained by the five-dimensional origin of the theory,
and are given by combinations of the four-dimensional Planck and Hubble scales.
This na\"{\i}vely determines the scale at which non-renormalizable
operators provide important corrections.
For example, when calculating the Newtonian potential, this reasoning suggests that
non-renormalizable operators become important
near energies~$\gtrsim 1/\ell$ \cite{Nicolis:2004qq},
where $\ell \sim 1000\;\mbox{km}$.
In principle
one should not trust computations of the classical gravitational
potential at distances $\lesssim \ell$.
However, it has long been known that an effect similar to the
Vainshtein mechanism suppresses these unwanted corrections,
making the theory predictive at least to centimetre scales.
In this section we will
demonstrate that our freedom to set the Lagrangian parameters
of the Galileon effective theory effectively means that
corrections can be made to enter at almost arbitrarily high scales.

For clarity
we denote fluctuations around the classical solution
by $\phi$, so that the full field can be written
$\pi_{\mathrm{dS}} + \phi$.
Making a field redefinition $\hat{\phi}=d_2^{1/2}\phi$,
the action for $\hat{\phi}$ can be written schematically
\begin{equation}
	\mathcal{L}_{\hat{\pi}}
	=
	- (\partial \hat{\phi}\cdot \partial \hat{\phi})
	\left(
		\frac{1}{2}
		+ \frac{1}{2} \frac{d_3}{d_2^{3/2}} [P]
		+ \frac{1}{4} \frac{d_4}{d_2^2} [P]^2
		+ \frac{1}{5} \frac{d_5}{d_2^{5/2}} [P]^3
	\right) ,
	\label{eq:qstrong}
\end{equation}
where all indices have been suppressed and $[P]$ indicates a series
of terms of
the form $\partial\partial\hat{\phi}$, with the indices variously
contracted. On inspection of this Lagrangian we
might expect strong interactions to enter
near any or all of the
energies $\sim d_2^{1/2}/d_3^{1/3}$, $\sim  d_2^{1/3}/d_4^{1/6}$ or
$\sim  d_2^{5/18}/d_5^{1/9}$.  

In the Galileon scenario these estimates can be rather misleading.
It was explained in \S\ref{sec:suppress} that
no gravitational experiment is sensitive to
fluctuations around a de Sitter background; instead, such experiments
measure fluctuations in the neighbourhood of massive objects.
Within the Vainshtein radius, renormalizations coming from the background
significantly alter the strong interaction scale.
For illustration, consider the case where
$c_4=c_5=d_4=d_5=0$. The 
Lagrangian for quantum fluctuations $\phi$ about the solution for the
$\pi$-field around a massive body is
\begin{equation}
\mathcal{L}_{\phi}=Z_{\mu\nu}\partial^{\mu}\phi\partial^{\nu}\phi-\frac{d_3}{2}(\Box\phi)\partial\phi\cdot\partial\phi ,
\end{equation}
where 
\begin{equation}
Z_{\mu\nu} =
-\frac{d_2}{2}\eta_{\mu\nu}-d_3\Box\pi\eta_{\mu\nu}+d_3\partial_{\mu}\partial_{\nu}\pi .
\end{equation}
Deep inside the Vainshtein radius of a massive object,
where $r\ll R_1$, all non-zero entries of $Z_{\mu\nu}$ become large
and of the same order.
We will discuss $Z_{\mu\nu}$ in more detail below.
In the present case we are assuming
$d_4=0$, and there is no Region B.
Therefore, within the Vainshtein radius,
the $r$-dependence $\pi$ is determined by the $\mathcal{L}_3$ operator,
and the non-zero elements of $Z_{\mu\nu}$ are of an approximately
common magnitude $Z$,
\begin{equation}
Z\sim
-\frac{d_2}{2}\left[1-\frac{1}{2}\left(\frac{R_1}{r}\right)^{3/2}\right]
.
\label{eq:Z}
\end{equation}
When $Z$ is large, this raises
the effective quantum cutoff to $\tilde{\Lambda}(x)\sim
Z^{1/2}/d_3^{1/3}$. 

The improvement is significant.
In computations of the Newtonian potential at the surface of the Earth,
this raises the cutoff in the DGP model to
$\tilde{\Lambda}(x)\sim
1/(1\;\mbox{cm})$ \cite{Nicolis:2004qq}.  Unfortunately,
this is two orders of magnitude lower than the scales probed in current
gravitational experiments, which can reach $\sim 10^{-2}\;\mbox{cm}$.
It is not yet clear whether this is in
conflict with observation.
In the Galileon model with $c_4=c_5=0$, the bound given in
Eq.~\eref{eq:constr} implies that we also find
\begin{equation}
\tilde{\Lambda}\lesssim \frac{1}{1 \mbox{ cm}} .
\end{equation}
Therefore, in this model, our freedom to change the Lagrangian parameters
does not significantly
improve our control over non-renormalizable operators.

The situation is different
in the full Galileon model with all the $c_i$ non-zero.
The possibility of a non-trivial Region B emerges and we have more
freedom. Within the Vainshtein radius,
the behaviour of $\pi$ as a function of $r$ depends whether
$\mathcal{L}_3$ or $\mathcal{L}_4$ dominates its dynamics.
Therefore, the strong coupling scale will vary between Region A and Region B.
It follows from the expressions given in Ref.~\cite{Nicolis:2008in}
that the quadratic Lagrangian for small perturbations
about $\pi$ satisfies
\begin{equation}
S_{\phi}=\frac{1}{2}\int d^4x\;[Z_t(r)(\partial_t\phi)^2-Z_r(r)(\partial_r\phi)^2-Z_{\Omega}(r)(\partial_{\Omega}\phi)^2]
\end{equation}
where $\partial_{\Omega}$ denotes derivatives with respect to
angular variables.  The $Z$ coefficients take different forms in the
two Regions A and B but, as with
the simple example discussed above, 
within the
Vainshtein radius they are always
 much larger than the na\"{\i}ve scale $d_2$.

\begin{itemize}
\item {\bf Region A:} The $Z$ coefficients scale as
\begin{eqnarray}
Z_r&\sim& 2d_2\left(\frac{R_1}{r}\right)^{3/2} ,\\
Z_t&\sim&\frac{3d_2}{2}\left(\frac{R_1}{r}\right)^{3/2}-\frac{3d_5d_2^3}{2d_3^3}\left(\frac{R_1}{r}\right)^{9/2}
,\\
Z_{\Omega}&\sim&  \frac{d_2}{2}\left(\frac{R_1}{r}\right)^{3/2}.
\end{eqnarray}
If the contribution proportional to $d_5$ is negligible,
then all of the
$Z_i$ have the same behaviour. This is analogous to the
simple example with $c_4=c_5=0$, discussed above.  The lower bound on
$d_3$ given by
Eq.~\eref{eq:constr} means that at fixed radius $r$, the
$Z_i$ are bounded above and cannot be made arbitrarily large.
Therefore, at that point the strong coupling scale cannot be pushed
to arbitrarily small distances.

\item {\bf Region B:}  The $Z$ coefficients scale as
\begin{eqnarray}
Z_r&\sim& 3d_2\left(\frac{R_2}{r}\right)^{2} ,\\
Z_t&\sim&3d_2\left(\frac{R_2}{r}\right)^{2}-\frac{d_5d_2^{3/2}}{\sqrt{2}d_4^{3/2}}\left(\frac{R_2}{r}\right)
,\\
Z_{\Omega}&\sim&  -d_2+\frac{2d_3^2}{3d_4} .
\end{eqnarray}
In this region the $Z_i$ scale differently with $r$. We see that
by varying the $d_i$ it is possible to make all of these coefficients
arbitrarily large.
\end{itemize}

In a
general Galileon theory, where all of the $c_i$ are non-zero,
not only $\mathcal{L}_2$ will be renormalized, but also
$\mathcal{L}_3$ and $\mathcal{L}_4$.
Therefore it is unclear which term in Eq.~\eref{eq:qstrong} gives
rise to the leading order strong-coupling corrections,
because we have no
understanding of the natural scales for each $c_i$.
For example, if the leading quantum corrections
come from the $\mathcal{L}_5$ operator, then the strong
coupling scale is $\sim Z(r)^{5/18}/d_5^{1/9}$. We have seen above
that the form of $Z(r)$ depends whether the point of
evaluation lies in Region A or Region B, but it is clear that in either
regime we are free to vary the Lagrangian parameters so that
the strong coupling scale lies beyond the reach of current experiments.

\section{Conclusions}
\label{sec:conc}

The Galileon is a scalar field coupled to the matter content of
the Standard Model. Its effects can be hidden from searches for
modifications of gravity, without gross
fine-tuning of parameters, because its
non-linear self-interactions cause the field to
become weakly coupled near heavy objects. The form of these self
interactions is protected by a derivative shift symmetry.

Our discussion has been framed in the context of an effective field theory,
where the Galileon's behaviour is controlled by a small number of
relevant couplings.
In some previous discussions, compatibility with a higher-dimensional
completion has fixed these couplings and therefore
the Vainshtein radius.
In Ref.~\cite{Nicolis:2008in} the couplings were allowed to vary
freely, but the far-field interaction strength, controlled by the
coupling $d_2$, was set to gravitational strength.
In this article we have shown that the Vainshtein radius
associated with
a massive object can be made arbitrarily large or small by varying
a linear combination of the
Lagrangian parameters.
Experiments probing the force on a test mass interacting
with a heavy source at fixed radius can be made insensitive to
Galileon exchange forces
by varying an independent linear combination of Lagrangian parameters.
In principle
this allows the far-field interaction strength to be made much
stronger. Whether such an arrangement can be technically natural
is presently unknown.

Calculation
of the precise constraints imposed by experimental tests of gravity
is left for future work. However, it seems reasonable
that observations do not force us to put all the Lagrangian
parameters at the Planck scale, as had been supposed.
Instead, they may vary widely over many orders of magnitude.  This is
particularly interesting in light of the recently proposed
five-dimensional origin of the Galileon model \cite{deRham:2010eu},
where the Galileon parameters derive from combinations of the
parameters of the five-dimensional theory.
The freedom to vary the Galileon parameters also means that
the quantum strong coupling scale can be made arbitrarily high, and
placed out of reach of current experimental probes.

\section*{Acknowledgments}
We would like to thank Claudia de Rham and Andrew Tolley for very
helpful discussions.  CB is supported by the German Science Foundation (DFG) under the
Collaborative Research Centre (SFB) 676, and would like to thank
the University of Sussex for their hospitality while this work was
completed. DS is supported by STFC.

\appendix

\section{Force on a test particle}

To describe the effects of the Galilean field on a test particle we
determine its geodesic equation, including the influence of
mixing between $\pi$ and the metric.
In the Galileon theory we treat the gravitational field,
and $\pi$'s contribution to it, at linear order, and measure length
along the worldline of a point particle using the metric
$g_{ab}=(1+2\pi)\eta_{ab}$, \cite{Nicolis:2008in}.
The action of a point particle of mass $m$ is 
\begin{equation}
	S = - m \int \d\tau \; \sqrt{1+2\pi}(-\dot{X}^a\dot{X}_a)^{1/2}
\end{equation} 
where the $X^a$ describe an embedding of the particle's worldline into
spacetime, parametrized by $\tau$.
Variation of this action leads us to the Galileon analogue of the
geodesic equation
\begin{equation}
\frac{ du^b}{d \tau} = -\nabla_a\pi
[\delta^a_b(-\dot{X}^c\dot{X}_c)^{1/2}+u^au^b]
\label{eq:force}
\end{equation}
where $u^a=\dot{X}^a/\sqrt{-\dot{X}^b\dot{X}_b}$.  In \S\ref{sec:fifth} we  considered only static, spherically
symmetric solutions for $\pi$.  For such solutions we see from
(\ref{eq:force}) that the force per unit mass on a point particle due
to the Galileon field is 
\begin{equation}
F_{\pi}= \frac{\d \pi(r)}{\d r}.
\end{equation}

\section*{References}

\bibliographystyle{JHEP}
\bibliography{galileon_paper}

\end{document}